\documentclass[prc,twocolumn,showpacs]{revtex4-1}

\usepackage{enumerate}

\usepackage{booktabs}
\usepackage{color}
\usepackage{graphicx}

\usepackage{amsmath}
\usepackage{amssymb}
\usepackage{times}

\usepackage{color}

\newcommand{\bra}{\langle}
\newcommand{\ket}{\rangle}

\newcommand{\beq}{\begin{equation}}
\newcommand{\eeq}{\end{equation}}
\newcommand{\bea}{\begin{eqnarray}}
\newcommand{\eea}{\end{eqnarray}}

\def\fun#1#2{\lower3.6pt\vbox{\baselineskip0pt\lineskip.9pt
 \ialign{$\mathsurround=0pt#1\hfil##\hfil$\crcr#2\crcr\sim\crcr}}}

\begin{document}

\title{
  Fano effect on the neutron elastic scattering by open-shell nuclei
}

\author{K. Mizuyama$^{1}$, N. Nhu Le$^{2}$, T. V. Nhan Hao$^{2}$}

\affiliation{
  \textsuperscript{1} Institute of Research and Development, Duy Tan University,  Da Nang 550000, Vietnam \\
  \textsuperscript{2} Faculty of Physics, University of Education, Hue University, 34 Le Loi Street, Hue City, Vietnam
}


\email{corresponding author: T. V. Nhan Hao (tvnhao@hueuni.edu.vn)}

\date{\today}

\begin{abstract}
  By focusing on the asymmetric shape of cross section, we analyze the pairing effect on the partial
  wave components of cross section for neutron elastic scattering off stable and unstable nuclei within
  the Hartree-Fock-Bogoliubov (HFB) framework. Explicit expressions for Fano parameters $q_{lj}$ and
  $\epsilon_{lj}$ have been derived and the pairing effects have been analyzed in term of these parameters,
  and the Fano effect was found on the neutron elastic scattering off the stable nucleus in terms of
  the pairing correlation. Fano effect was appeared as the asymmetric line-shape of the cross section
  caused by the small absolute value of $q_{lj}$ due to the small pairing effect on the deep-lying
  hole state of the stable nucleus. In the case of the unstable nuclei, the large $q_{lj}$ value
  is expected because of the small absolute value of the Fermi energy. The quasiparticle resonance
  with the large $q_{lj}$ forms the Breit-Wigner type shape in the elastic scattering cross section.
\end{abstract}

\maketitle

\section{Introduction}
The Fano effect~\cite{fano} has been known as an universal quantum phenomenon
in which the transition probability becomes a characteristic asymmetric shape
caused by the interference effect due to the correlation between discrete
states (or resonance) and continuum.
Many examples of Fano effect can be found
in physics even though the mechanism is quite different for each example.
The Raman scattering~\cite{raman,raman2}, the photoelectric
emission~\cite{photoe}, photoionization~\cite{atom},
the photoabsorption~\cite{photoabs} and the neutron scattering~\cite{nscat}
are examples which have been known in spectroscopy.
Recently, in atomic and condensed matter physics, experimental research to
control the Fano effect has been started~\cite{ott, ryu} in order to investigate
the detailed dynamical system of the Fano effect.
Also in nuclear physics, some observed resonances have been reported as
the candidate of the Fano resonance~\cite{orrigo, orrigo2}, however, there was no
detailed study/analysis in terms of the Fano parameters, so far.

In nuclear physics, sharp resonances found in the experimental data of
$^{15}$N($^{7}$Li,$^{7}$Be)$^{15}$C reaction have been analyzed by using the
channel-coupling equation, and introduced as the candidates for the Fano
resonance~\cite{orrigo}.
Another candidate for the Fano resonance in nuclei is the quasiparticle
resonance (or pair resonance) due to the particle-hole configuration
mixing caused by the pairing effect at the ground state of the open
shell nuclei. The experimental cross section data of d($^{9}$Li,$^{10}$Li)p
reaction has been analyzed in terms of the effect of the pair resonance
basing on the HFB formalism~\cite{orrigo2}.
Despite mentioning the possibility that these two candidates
introduced in previous studies are Fano effect, there was no
detailed study on Fano parameters.

The aim of this study is to organize the Fano formula for the neutron
elastic scattering on the open shell nuclei with the help of the
Jost function formalism~\cite{jost} based on the HFB~\cite{jost-hfb}, and
we shall discuss the role of the Fano parameters for the quasiparticle resonances
seen in the partial cross section of the neutron elastic scattering.
Because characteristic asymmetric shapes of the
cross sections have been shown as the numerical results obtained by
the HFB framework~\cite{jost-hfb}.

Firstly, we divide the Jost function into two parts: the scattering
part and the pairing part. The Gell-Mann-Goldberger relation
for the T-matrix (expressed by the Jost function) has been obtained based on the HFB approach.
Secondly, we derive the explicit expressions of the Fano parameters $q$
and $\epsilon$ for the neutron elastic scattering by open-shell nuclei within
the HFB framework.
Finally, the role of the Fano parameters has been analyzed in comparison with the
square of the T-matrix plotted as a function of the incident energy of neutron.

\section{Method}

\subsection{Gell-Mann-Goldberger relation in HFB}
The Jost function based on the HFB~\cite{jost-hfb}
can be divided as
\begin{eqnarray}
  &&
  \left(
  \mathcal{J}_{lj}^{(\pm)}(E)
  \right)_{s1}
  \nonumber\\
  &&=
  \delta_{s1}
  J_{0,lj}^{(\pm)}(k_1(E))
  \nonumber\\
  &&
  \mp
  \frac{2m}{\hbar^2}
  \frac{k_1(E)}{i}
  \int dr
  \varphi_{0,lj}^{(\pm)}(r;k_1(E))
  \Delta(r)
  \varphi_{2,lj}^{(rs)}(r;E),
  \label{eq1}
\end{eqnarray}
using the Hartree-Fock (HF) solutions
$\varphi_{0,lj}^{(\pm)}(r;k_1(E))$ which satisfy
the out-going/in-coming boundary conditions, and $\varphi_{2,lj}^{(rs)}(r;E)$ is the
lower component of the HFB solution which is regular at the origin $r=0$, where
$E$ is the quasiparticle energy and $k_1(E)$ is the momentum defined by
$k_1(E)=\sqrt{\frac{2m}{\hbar^2}(\lambda+E)}$ with the Fermi energy $\lambda (<0)$.
$J_{0,lj}^{(\pm)}(k_1(E))$ is the HF Jost function given by
\begin{eqnarray}
  &&
  J_{0,lj}^{(\pm)}(k_1(E))
  \nonumber\\
  &&=
  1
  \nonumber\\
  &&
  \mp
  \frac{2m}{\hbar^2}
  \frac{k_1(E)}{i}
  \int dr
  rh^{(\pm)}_l(k_1(E)r)
  U_{lj}(r)
  \varphi_{0,lj}^{r}(r;k_1(E)),
  \nonumber\\
  \label{eq2}
\end{eqnarray}
$U_{lj}(r)$ and $\Delta(r)$ are the HF mean field and the pair potential,
respectively. We adopt the same Woods-Saxon form and their parameters
as \cite{jost-hfb} for the numerical calculation. $\varphi_{0,lj}^{r}(r;k_1(E))$
is the regular solution of the HF equation. $\varphi_{0,lj}^{r}(r;k_1(E))$ and
$\varphi_{0,lj}^{(\pm)}(r;k_1(E))$ are connected by using $J_{0,lj}^{(\pm)}(k_1(E))$ as
\begin{eqnarray}
  &&
  \varphi_{0,lj}^{r}(r;k_1(E))
  =
  \frac{1}{2}
  \left[
    J_{0,lj}^{(+)}(k_1(E))\varphi_{0,lj}^{(-)}(r;k_1(E))
    \right.
  \nonumber\\
  &&\hspace{1cm}
  \left.
  +
  J_{0,lj}^{(-)}(k_1(E))\varphi_{0,lj}^{(+)}(r;k_1(E))
  \right].
  \label{eq3}
\end{eqnarray}
From Eqs.(\ref{eq1}) and (\ref{eq3}), we derive
\begin{eqnarray}
  &&
  \left(
  \mathcal{J}_{lj}^{(-)}(E)
  \right)_{s1}
  J_{0,lj}^{(+)}(k_1(E))
  -
  \left(
  \mathcal{J}_{lj}^{(+)}(E)
  \right)_{s1}
  J_{0,lj}^{(-)}(k_1(E))
  \nonumber\\
  &&=
  \frac{2m}{\hbar^2}
  \frac{2k_1(E)}{i}
  \int dr
  \varphi_{0,lj}^{r}(r;k_1(E))
  \Delta(r)
  \varphi_{2,lj}^{(rs)}(r;E).
  \label{eq4}
\end{eqnarray}

Applying the HFB T-matrix given by Eq.~(60) in ~\cite{jost-hfb}
and the HF T-matrix given by
\begin{eqnarray}
  &&
  T_{lj}^{(0)}(E)
  =
  \frac{i}{2}
  \left(
  \frac{J_{0,lj}^{(-)}(k_1(E))}{J_{0,lj}^{(+)}(k_1(E))}
  -1
  \right)
  \nonumber\\
  &&=
  \frac{2mk_{1}(E)}{\hbar^2}
  \int_0^\infty
  dr
  rj_l(k_1(E)r)
  U_{lj}(r)
  \psi_{0,lj}^{(+)}(r;k_1(E))
  \nonumber\\
  &&=
  \frac{2mk_1(E)}{\hbar^2}
  \bra j_l(k_1(E))|U_{lj}|\psi_{0,lj}^{(+)}(k_1(E))\ket.
  \label{T0int}
\end{eqnarray}
to Eq.(\ref{eq4}),
we obtain
\begin{eqnarray}
  &&
  T_{lj}(E)
  -
  T_{lj}^{(0)}(E)
  \nonumber\\
  &&=
  \frac{2mk_1(E)}{\hbar^2}
  \int dr
  \psi_{0,lj}^{(+)}(r;k_1(E))
  \Delta(r)
  \psi_{2,lj}^{(+)}(r;E),
  \label{eq6}
\end{eqnarray}
where $\psi_{0,lj}^{(\pm)}(r;k_1(E))=\varphi_{0,lj}^{r}(r;k_1(E))/J_{0,lj}^{(\pm)}(k_1(E))$
and $\psi_{2,lj}^{(+)}(r;E)$ is the lower component of the HFB scattering wave
function~\cite{jost-hfb}.

Eq.~(\ref{eq6}) is the so-called ``Gell-Mann-Goldberger relation''
(two potential formula)~\cite{gellmann,Hussein} in the HFB formalism.
We can read that the right hand side of Eq.~(\ref{eq6}) represents the transition from the
hole-like component (lower component) of the HFB scattering states to the
HF scattering states caused by the pairing field $\Delta(r)$.
The HFB scattering wave function
$\psi_{lj}^{(+)}(r;E)=
\begin{pmatrix}
  \psi_{1,lj}^{(+)}(r;E) \\
  \psi_{2,lj}^{(+)}(r;E)
\end{pmatrix}$
can be represented in the integral form as
\begin{eqnarray}
  &&
  \begin{pmatrix}
    \psi_{1,lj}^{(+)}(r;E) \\
    \psi_{2,lj}^{(+)}(r;E)
  \end{pmatrix}
  =
  \begin{pmatrix}
    \psi_{0,lj}^{(+)}(r;k_1(E)) \\
    0
  \end{pmatrix}
  \nonumber\\
  &&
  +
  \int dr'
  \begin{pmatrix}
    \mathcal{G}_{lj}^{11}(r,r';E) & \mathcal{G}_{lj}^{12}(r,r';E) \\
    \mathcal{G}_{lj}^{21}(r,r';E) & \mathcal{G}_{lj}^{22}(r,r';E)
  \end{pmatrix}
  \nonumber\\
  &&\hspace{10pt}
  \times
  \begin{pmatrix}
    0 & \Delta(r') \\
    \Delta(r') & 0
  \end{pmatrix}
  \begin{pmatrix}
    \psi_{0,lj}^{(+)}(r';k_1(E)) \\
    0
  \end{pmatrix},
  \label{psiint}
\end{eqnarray}
using the HFB Green's function given by $2\times 2$ matrix form~\cite{fayans,matsuo,belyaev}.

Inserting Eq. (\ref{psiint}) into the r.h.s. of Eq. (\ref{eq6}), one obtains
\begin{eqnarray}
  &&
  T_{lj}(E)
  -
  T_{lj}^{(0)}(E)
  \nonumber\\
  &&
  =
  \frac{2mk_1(E)}{\hbar^2}
  \int\int drdr'
  \nonumber\\
  &&\times
  \psi_{0,lj}^{(+)}(r;k_1(E))
  \Delta(r)
  \mathcal{G}_{lj}^{22}(r,r';E)
  \Delta(r')
  \psi_{0,lj}^{(+)}(r';k_1(E)).
  \nonumber\\
  \label{eq7}
\end{eqnarray}

Since it has been proved that $S_{lj}^{11}(E)$ satisfies the unitarity on the scattering states defined on
the real axis of $E$ above the Fermi energy $-\lambda$ in ~\cite{jost-hfb},
$S_{lj}^{11}(E)$ can be expressed as $S_{lj}^{11}(E)=e^{2i\delta_{lj}(E)}$.
(The quasiparticle energy $E$ is hereafter supposed to be on the scattering states.)
Also $S_{lj}^{(0)}(E)$
can be expressed $S_{lj}^{(0)}(E)=e^{2i\delta_{lj}^{(0)}(E)}$ because of no absorption in $U_{lj}(r)$.
Here, let us define a phase-shift as $\delta_{lj}^{(1)}\equiv\delta_{lj}-\delta_{lj}^{(0)}$ to
define $T_{lj}^{(1)}\equiv -e^{i\delta_{lj}^{(1)}(E)}\sin\delta_{lj}^{(1)}(E)$, and
it may be rather trivial that $T_{lj}(E)$, $T_{lj}^{(0)}(E)$ and $T_{lj}^{(1)}(E)$ are related by
\begin{eqnarray}
  T_{lj}(E)
  &=&
  T_{lj}^{(0)}(E)
  +
  T_{lj}^{(1)}(E)
  S_{lj}^{(0)}(E)
  \label{T1}\\
  &=&
  T_{lj}^{(0)}(E)
  \left(
  1-iT_{lj}^{(1)}(E)
  \right)
  \nonumber\\
  &&
  +
  T_{lj}^{(1)}(E)
  \left(
  1-iT_{lj}^{(0)}(E)
  \right)
  \label{T1-2}.
\end{eqnarray}

\subsection{Fano parameters}
In this paper, we analyze the pairing effect on the partial
cross sections of $p_{1/2}$ with $\lambda=-8.0$ MeV and $d_{3/2}$ with
$\lambda=-1.0$ MeV by using the same Woods-Saxon parameters for the numerical
calculation as in Ref. ~\cite{jost-hfb}.
Both resonances are the so-called {\it hole-type}
resonances originated from the hole state resulting from the particle-hole (p-h)
configuration mixing due to the pairing. As shown in Fig.6 of \cite{jost-hfb},
there is only one hole state for both $p_{1/2}$ and $d_{3/2}$ at the no pairing
limit.

In such cases, the Hartree-Fock Green function can be expressed as
\begin{eqnarray}
  &&
  G_{HF,lj}(r,r';\epsilon(k))
  \nonumber\\
  &&
  =
  \frac{\phi_{h,lj}(r)\phi^*_{h,lj}(r')}{\epsilon(k)-e_h}
  -i
  \frac{2mk}{\hbar^2}
  \psi_{0,lj}^{(+)}(r;k)\psi_{0,lj}^{(+)*}(r';k)
  \nonumber\\
  &&
  +
  \frac{2m}{\hbar^2}
  \frac{2}{\pi}
  P
  \int_0^\infty dk'k'^2
  \frac{\psi_{0,lj}^{(+)}(r;k')\psi_{0,lj}^{(+)*}(r';k')}{k^2-k'^2},
  \label{GHFsp}
\end{eqnarray}
by dividing the continuum part into the principal and other parts
in the spectral representation. Here $\epsilon(k)=\frac{\hbar^2k^2}{2m}$.

Using Eq.(\ref{GHFsp}), we obtain
\begin{eqnarray}
  &&
  \bra \phi_{h,lj}|\mathcal{G}_{lj}^{22}(E)|\phi_{h,lj}\ket
  =
  \frac{1}{E-\lambda+e_h-F_{lj}(E)+i\Gamma_{lj}(E)/2},
  \label{g22-hh}
  \nonumber\\
  \\
  &&
  F_{lj}(E)
  =
  \frac{2m}{\hbar^2}
  \frac{2}{\pi}
  P
  \int_0^\infty dk'k'^2
  \frac{|\bra \psi_{0,lj}^{(+)}(k')|\Delta|\phi_{h,lj}\ket|^2}{k^2_1(E)-k'^2},
  \label{defF}
  \\
  &&
  \Gamma_{lj}(E)/2
  =
  \frac{2m k_1(E)}{\hbar^2}
  |\bra \psi_{0,lj}^{(+)}(k_1(E))|\Delta|\phi_{h,lj}\ket|^2,
  \label{defgm}
\end{eqnarray}
as an exact solution of the HFB Dyson equation for $\mathcal{G}_{lj}$.

Using Eqs.~(\ref{eq7}), (\ref{T1-2}) and (\ref{g22-hh}), we derive
\begin{eqnarray}
    T_{lj}^{(1)}(E)
    &=&
    \frac{\Gamma_{lj}(E)/2}{E-\lambda+e_h-F_{lj}(E)+i\Gamma_{lj}(E)/2}
    \label{T1-3}.
\end{eqnarray}
This is the typical Breit-Wigner formula for
the hole-type quasiparticle resonance.
From this formula, we can notice that the hole state which satisfies
$2\lambda-e_h-F_{lj}(E) >0$
can be observed as the quasiparticle resonance in the neutron elastic
scattering cross section since the incident neutron energy $E_i$ is defined
by $E_i=E+\lambda$.

One of the parameters introduced by U.Fano \cite{fano}, $\epsilon_{lj}(E)$
is defined by
\begin{eqnarray}
  \epsilon_{lj}(E)
  &=&
  \frac{1-iT_{lj}^{(1)}(E)}{T_{lj}^{(1)}(E)}
  \label{edef1}
  \\
  &=&
  \frac{E-\lambda+e_h-F_{lj}(E)}{\Gamma_{lj}(E)/2}.
  \label{edef2}
\end{eqnarray}
We notice that the quasiparticle
resonance energy $E_r$ and the width $\Gamma_{lj}(E_r)$ can be estimated as
\begin{eqnarray}
  \epsilon_{lj}(E=E_r)&=&0,
  \label{Er-epsilon}
  \\
  \left.\frac{d\epsilon_{lj}(E)}{dE}\right|_{E=E_r}
  &=&
  \frac{2}{\Gamma_{lj}(E_r)},
  \label{gamma-epsilon}
\end{eqnarray}
by using Eq.(\ref{edef2}).

When $\bra \psi_{0,lj}^{(+)}(k_1(E))|\Delta|\phi_{h,lj}\ket\neq 0$,
we can obtain
\begin{eqnarray}
  &&
  1-iT_{lj}^{(0)}(E)
  \nonumber\\
  &&=
  \frac{\bra\psi_{0,lj}^{(+)}(k_1(E))|\Delta|\phi_{h,lj}\ket
    -iT_{lj}^{(0)}(E)\bra\psi_{0,lj}^{(+)}(k_1(E))|\Delta|\phi_{h,lj}\ket}
    {\bra\psi_{0,lj}^{(+)}(k_1(E))|\Delta|\phi_{h,lj}\ket}
  \nonumber\\
  &&=
  \frac{
    \bra\psi_{0,lj}^{(+)}(k_1(E))|\Delta|\phi_{h,lj}\ket
    -
    \bra j_l(k_1(E))|U_{lj}G^*_{HF,lj}\Delta|\phi_{h,lj}\ket}{%
    \bra\psi_{0,lj}^{(+)}(k_1(E))|\Delta|\phi_{h,lj}\ket}
  \nonumber\\
  &&
  \hspace{10pt}
  +
  \frac{2m}{\hbar^2}
  \frac{2}{\pi}
  P
  \int_0^\infty dk'k'^2
  \nonumber\\
  &&\hspace{20pt}
  \times
  \frac{\bra j_l(k_1(E))|U_{lj}|\psi_{0,lj}^{(+)}(k')\ket
    \bra\psi_{0,lj}^{(+)}(k')|\Delta|\phi_{h,lj}\ket}{%
    (k_1^2(E)-k'^2)\bra\psi_{0,lj}^{(+)}(k_1(E))|\Delta|\phi_{h,lj}\ket},
  \nonumber\\
  \label{1T0}
\end{eqnarray}
by using Eqs.(\ref{T0int}) and (\ref{GHFsp}).
Note that $\bra j_l(k_1(E))|U_{lj}|\phi_{h,lj}\ket=0$ is also used.

\begin{widetext}
Since $|\psi_{0,lj}^{(+)}(k)\ket$ satisfies the Lippmann-Schwinger
equation $|\psi_{0,lj}^{(+)}(k)\ket=(1+G_{HF,lj}U_{lj})|j_l(k)\ket$, we
rewrite Eq.(\ref{1T0}) as
\begin{eqnarray}
  1-iT_{lj}^{(0)}(E)
  =
  \frac{
  \bra j_l(k_1(E))|\Delta|\phi_{h,lj}\ket
  +
  \displaystyle{
  \frac{2m}{\hbar^2}
  \frac{2}{\pi}
  P
  \int_0^\infty dk'k'^2
  \frac{
    \bra j_l(k_1(E))|U_{lj}|\psi_{0,lj}^{(+)}(k')\ket
    \bra\psi_{0,lj}^{(+)}(k')|\Delta|\phi_{h,lj}\ket
  }{%
    k_1^2(E)-k'^2
  }}
  }{%
    \bra\psi_{0,lj}^{(+)}(k_1(E))|\Delta|\phi_{h,lj}\ket
  }.
  \label{1T0-2}
\end{eqnarray}
\end{widetext}
Using Eq.(\ref{1T0-2}), another parameter $q_{lj}(E)$ is defined by
\begin{eqnarray}
  q_{lj}(E)
  &=&
  \frac{1-iT_{lj}^{(0)}(E)}{T_{lj}^{(0)}(E)}
  \label{qdef1}\\
  &&
  =
  \frac{
  \bra \chi_l(k_1(E))|\mathcal{U}_{lj}|\Phi_{h,lj}\ket
  }{%
    T_{lj}^{(0)}(E)
    \bra\psi_{0,lj}^{(+)}(k_1(E))|\Delta|\phi_{h,lj}\ket
  },
  \label{qdef2}
\end{eqnarray}
where
\begin{eqnarray}
  &&
  |\Phi_{h,lj}\ket
  =
  \begin{pmatrix}
    |\Phi_{h,lj}^{(1)}\ket \\
    |\Phi_{h,lj}^{(2)}\ket
  \end{pmatrix}
  \nonumber\\
  &&=
  \begin{pmatrix}
    \frac{2m}{\hbar^2}
  \frac{2}{\pi}
  P
  \int_0^\infty dk'k'^2
  \frac{
    |\psi_{0,lj}^{(+)}(k')\ket
    \bra\psi_{0,lj}^{(+)}(k')|\Delta|\phi_{h,lj}\ket
  }{%
    k_1^2(E)-k'^2
  } \\
    |\phi_{h,lj}\ket
  \end{pmatrix},
  \\
  &&
  |\chi_{l}(k)\ket
  =
  \begin{pmatrix}
    |j_l(k)\ket \\
    0
  \end{pmatrix},
  \hspace{10pt}
  \mathcal{U}_{lj}
  =
  \begin{pmatrix}
    U_{lj} & \Delta \\
    \Delta & -U_{lj}
  \end{pmatrix}.
\end{eqnarray}
The upper component of $|\Phi_{h,lj}\ket$ is originated from the admixture of a hole state and
continuum due to the pairing. Note that, Eq.~(\ref{qdef2}) is quite analogous to Eq.~(20)
in \cite{fano}.

\begin{figure}[htbp]
\includegraphics[scale=1.9,angle=0]{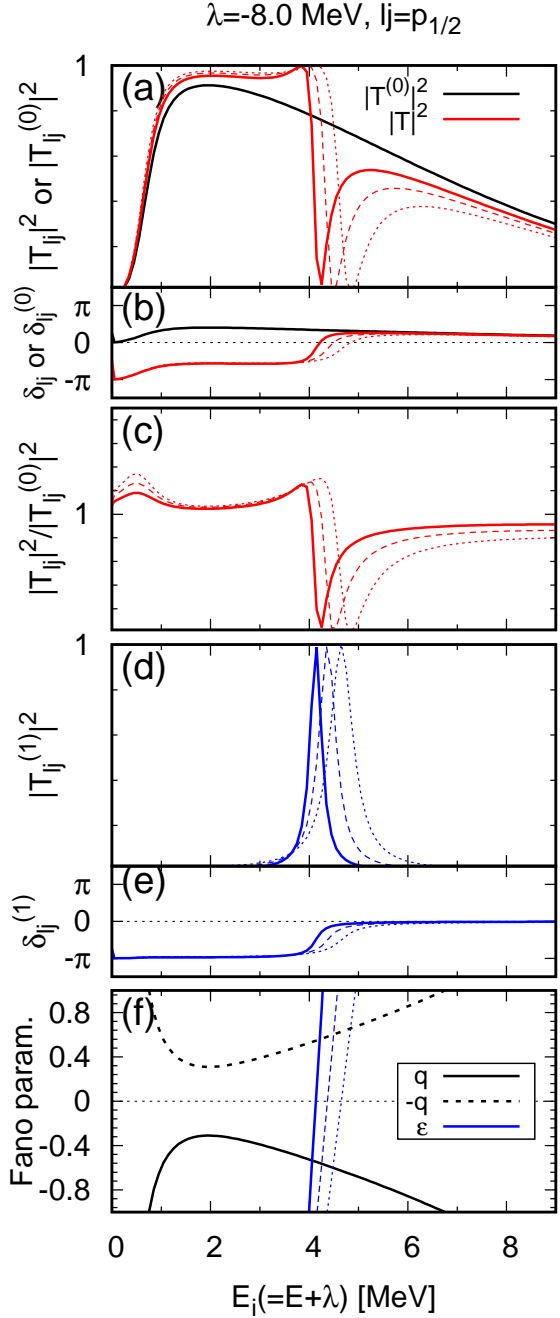}
\caption{(Color online) The numerical results of square of T-matrix of the neutron elastic scattering,
  corresponding phase-shift, and the Fano parameters $q$ and $\epsilon$ for $p_{1/2}$ plotted as a
  function of the incident neutron energy $E_i (=E+\lambda)$ with $\lambda=-8.0$ MeV. See text for details.}
\label{fig1}
\end{figure}
\begin{figure}[htbp]
\includegraphics[scale=1.9,angle=0]{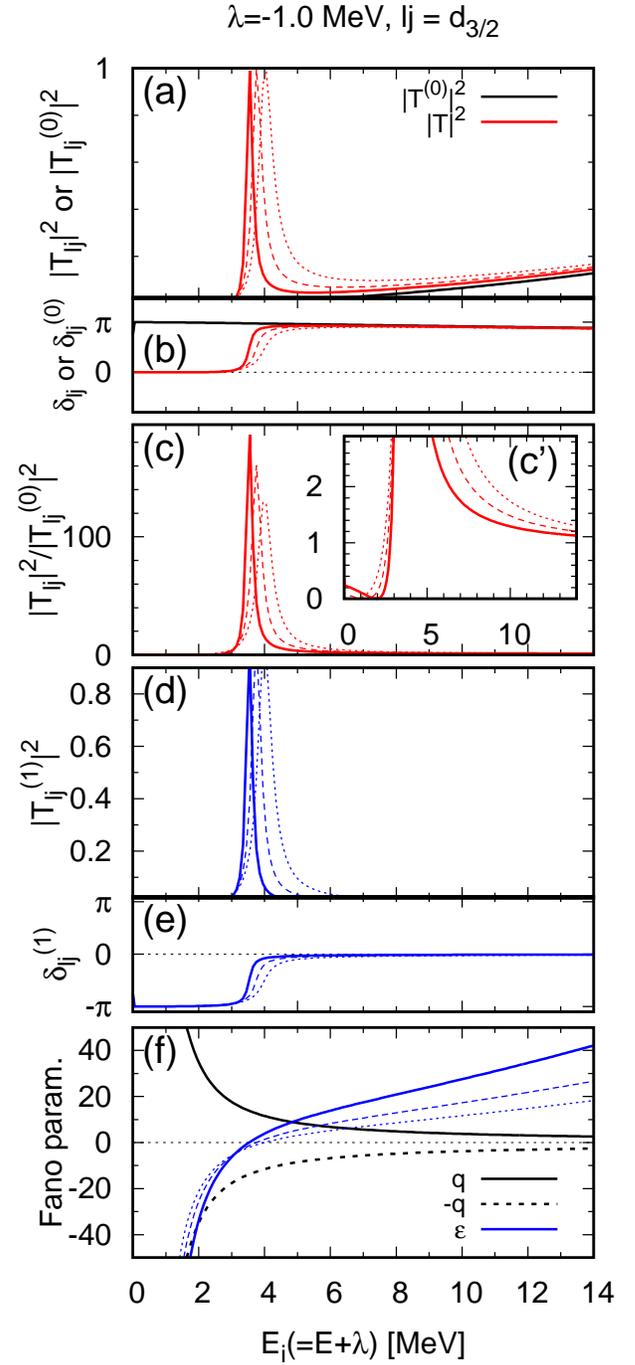}
\caption{(Color online) Same with Fig.\ref{fig1} but for $d_{3/2}$ with $\lambda=-1.0$ MeV.}
\label{fig2}
\end{figure}

\subsection{Fano formula}
Applying Eqs.~(\ref{edef1}) and (\ref{qdef1}) to Eq.(\ref{T1-2}),
it is rather easy to obtain
\begin{eqnarray}
  T_{lj}(E)
  &=&
  -
  e^{i\delta_{lj}^{(1)}(E)}
  \frac{
  q_{lj}(E)
  +
  \epsilon_{lj}(E)
  }{\sqrt{1+\epsilon^2_{lj}(E)}}
  T_{lj}^{(0)}(E).
\end{eqnarray}
Thus, we finally obtain the so-called {\it Fano} formula
\begin{eqnarray}
  \frac{|T_{lj}(E)|^2}{|T_{lj}^{(0)}(E)|^2}
  &=&
  \frac{
  (q_{lj}(E)
  +
  \epsilon_{lj}(E))^2
  }{1+\epsilon^2_{lj}(E)}.
  \label{fanoformula}
\end{eqnarray}
Using Eqs.~(\ref{qdef2}) and (\ref{defgm}), we obtain a
very similar formula to Eq.~(22) of Ref.~\cite{fano}
\begin{eqnarray}
  &&
  q_{lj}^2(E)/2
  =
  \frac{\hbar^2k_1^2(E)}{2m}
  \frac{
    |\frac{2m}{\hbar^2}\frac{1}{\sqrt{k_1(E)}}
    \bra \chi_l(k_1(E))|\mathcal{U}_{lj}|\Phi_{h,lj}\ket|^2
  }{%
    |T_{lj}^{(0)}(E)|^2
    \Gamma_{lj}(E)
  }
  \nonumber\\
  \label{q2}.
\end{eqnarray}
Since the resonance energy $E_r$ is determined by Eq.~(\ref{Er-epsilon}) and
$\Gamma_{lj}(E_r)$ is the width of the quasiparticle resonance, we obtain
\begin{eqnarray}
  &&
  \frac{|T_{lj}(E_r)|^2}{|T_{lj}^{(0)}(E_r)|^2}
  =
  q_{lj}^2(E_r)
  \\
  &&=
  \frac{
    |\frac{2m}{\hbar^2}\frac{1}{\sqrt{k_1(E_r)}}
    \bra \chi_l(k_1(E_r))|\mathcal{U}_{lj}|\Phi_{h,lj}\ket|^2
  }{%
    |T_{lj}^{(0)}(E_r)|^2
    \Gamma_{lj}(E_r)/2E_i^r
  }\label{q2er},
\end{eqnarray}
where $E_i^r=E_r+\lambda=\frac{\hbar^2k_1^2(E_r)}{2m}$.
The numerator of Eq.(\ref{q2er}) is the transition probability to the
``{\it modified quasi-hole}'' state at the resonance $E_r$.

Therefore, $q_{lj}^2(E_r)/2$ is regarded as the ratio of the transition
probabilities to the ``{\it modified quasi-hole}'' state $|\Phi_{h,lj}\ket$ and
to a scaled width $\Gamma_{lj}(E_r)/E_i^r$ of the HF continuum states
$|\psi_{0,lj}^{(+)}\ket$ at a quasiparticle resonance energy $E_r$.

As well known, the characteristic features of the Fano formula Eq.~(\ref{fanoformula}) are
\begin{enumerate}
\item
  The shape of $\left|\frac{T_{lj}(E)}{T_{lj}^{(0)}(E)}\right|^2$
  approaches the Breit-Wigner shape at the limit
  $|q_{lj}(E)|\to\infty$.
\item
  $\left|\frac{T_{lj}(E)}{T_{lj}^{(0)}(E)}\right|^2$ becomes zero at the energy
  $E=E_c$ which satisfies $q_{lj}(E_c)=-\epsilon_{lj}(E_c)$.
\end{enumerate}
Also the parameter $q_{lj}(E)$ causes the asymmetric shape of $\left|\frac{T_{lj}(E)}{T_{lj}^{(0)}(E)}\right|^2$
in $E$ as shown in Fig.1 of \cite{fano} when the absolute value of $q_{lj}(E)$ is non-zero small value
(see Fig. 1 of \cite{fano}),
$q_{lj}(E)$ is, therefore,  called {\it ``Fano asymmetry parameter''}~\cite{asymfano}.
Besides, it is clear from Eq.(\ref{T1-3}) that $|T_{lj}^{(1)}(E)|^2$ always keeps the shape of the
Breit-Wigner formula if a quasiparticle resonance exists.

\begin{table}
  \caption{%
    The pairing gap dependence of $E_r$, $\Gamma_{lj}(E_r)$, $q_{lj}(E_r)$ and $E_c$ for $p_{1/2}$
    with $\lambda=-8.0$ MeV and $d_{3/2}$ with $\lambda=-1.0$ MeV.
    The energies $E_r^i$ and $E_c^i$ defined by $E_i^r=E_r+\lambda$ and $E_i^c=E_c+\lambda$
    are shown in parenthesis.
    The results are shown in units of MeV except $q_{lj}$, $q_{lj}$ is a dimensionless quantity.
  }
  \label{table1}
  \begin{ruledtabular}
    \begin{tabular}{c|cccc|cccc}
      & \multicolumn{4}{c|}{$p_{1/2}$} & \multicolumn{4}{c}{$d_{3/2}$} \\
      & \multicolumn{4}{c|}{($\lambda=-8.0$ MeV)} & \multicolumn{4}{c}{($\lambda=-1.0$ MeV)} \\
      \colrule
      $\bra\Delta\ket$ &
      $E_r$ & $\Gamma_{lj}$ & $q_{lj}$ & $E_c$  &
      $E_r$ & $\Gamma_{lj}$ & $q_{lj}$ & $E_c$ \\
      & $(E_i^r)$ & & & $(E_i^c)$ & $(E_i^r)$ & & & $(E_i^c)$ \\
      \colrule
      $2.0$ & $12.14$  & $0.28$ & $-0.55$ & $12.21$  & $4.53$   & $0.20$ & $13.80$ & $2.94$   \\
            & $(4.14)$ &        &         & $(4.21)$ & $(3.53)$ &        &         & $(1.94)$ \\
      $2.5$ & $12.37$  & $0.42$ & $-0.58$ & $12.49$  & $4.75$   & $0.34$ & $12.67$ & $2.27$   \\
            & $(4.37)$ &        &         & $(4.49)$ & $(3.75)$ &        &         & $(1.27)$ \\
      $3.0$ & $12.65$  & $0.56$ & $-0.63$ & $12.84$  & $4.99$   & $0.52$ & $11.55$ & $1.45$   \\
            & $(4.65)$ &        &         & $(4.84)$ & $(3.99)$ &        &         & $(0.45)$
    \end{tabular}
  \end{ruledtabular}
\end{table}

\section{Numerical analysis}
Numerical results for $p_{1/2}$ with $\lambda=-8.0$ MeV and $d_{3/2}$ with $\lambda=-1.0$ MeV are shown
in Figs.\ref{fig1} and \ref{fig2}, respectively, by adopting the Woods-Saxon potential for the
mean field potential and pair potential with same parameters as \cite{jost-hfb}.

In panel (a) of Figs.~\ref{fig1} and \ref{fig2}, the square of T-matrix $|T_{lj}|^2$ and
$|T_{lj}^{(0)}|^2$ of the neutron elastic scattering are plotted as a function of the incident
neutron energy $E_i(=E+\lambda)$ by red curves and black curve. The solid red curve shows
$|T_{lj}|^2$ with $\bra\Delta\ket=2.0$ MeV. The dashed and dotted curves are the same ones with
$\bra\Delta\ket=2.5$ and $3.0$ MeV, respectively. Corresponding phase shifts are shown
in the panel (b). In panel (c), $|T_{lj}|^2/|T_{lj}^{(0)}|^2$ which is representative of the quantity of
the Fano formula Eq.(\ref{fanoformula}). We show $|T_{lj}^{(1)}|$ represented by Eq.(\ref{T1-3}) in the
panel (d), and the corresponding phase shifts $\delta_{lj}^{(1)}$ are shown in the panel (e).
The Fano parameters $q_{lj}$ and $\epsilon_{lj}$ are plotted by the solid black curve and the red curves
in the panel (f). The dashed black curve represents $-q_{lj}$. The pairing dependence for all quantities
in panels (b)-(f) is shown by solid, dashed and dotted curves (corresponding to
$\bra\Delta\ket=2.0, 2.5$ and $3.0$ MeV) as well as panel (a).
The values of $E_r$, $\Gamma_{lj}(E_r)$, $q_{lj}(E_r)$ and
$E_c$ for $p_{1/2}$ with $\lambda=-8.0$ MeV and $d_{3/2}$ with $\lambda=-1.0$ MeV are
shown in Table \ref{table1}. It is confirmed that these values of $E_r$ and $\Gamma_{lj}(E_r)$
matches with the zeros of the absolute values of the Jost function on the complex energy plane
shown in \cite{jost-hfb}.

In Fig.~\ref{fig1}, there is a dip at $E_i^c(=E_c+\lambda)$ and asymmetric shape
in $|T_{lj}|^2$(panel (a)) and $|T_{lj}|^2/|T_{lj}^{(0)}|^2$ (panel (c)).
According to the characteristic of the Fano formula Eq.(\ref{fanoformula}), this
asymmetric shape is due to the small absolute value of $q_{lj}$ at the resonance
energy $E_r$ as shown in panel (f) and Table \ref{table1}.
In the panel (d), $|T_{lj}^{(1)}|^2$ shows a typical Breit-Wigner resonance shape
representing a hole-type quasiparticle resonance originates from a deep-lying
hole state ($e_{p_{1/2}}=-19.71$ MeV) due to the pairing correlation.
The typical behaviour of the phase shift for a resonance can be seen in
the panel (e).
In Fig.~\ref{fig2}, there is a sharp resonance in the panels (a), (c) and (d).
This resonance originates from a hole state at $e_{d_{3/2}}=-5.12$ MeV.
According to the characteristic of the Fano formula Eq.~(\ref{fanoformula}), this
is due to the large absolute value of $q_{lj}$ at the resonance energy $E_r$
as shown in a panel (f) and Table \ref{table1}.
The typical behaviour of the phase shift for a resonance can be seen in
both panels (b) and (e).

In Table \ref{table1}, one can see that the energies $E_r$ and $E_c$ are shifted
to higher energy, and the width $\Gamma_{lj}(E_r)$ becomes larger as the pairing
gap $\bra\Delta\ket$ increases. These are rather trivial pairing effects, because the energy
shift $F_{lj}$ and width $\Gamma_{lj}$ are represented by Eqs.(\ref{defF}) and
(\ref{defgm}), and the pairing does not change the relative position between
$E_r$ and $E_c$.
However, the pairing effect on $q_{lj}(E_r)$ is not so simple. In the case of
$p_{1/2}$ with $\lambda=-8.0$ MeV, the absolute value of $q_{lj}(E_r)$ increases
as the pairing gap $\bra\Delta\ket$ increases. On the other hand, the absolute value of $q_{lj}(E_r)$
decreases as the pairing gap $\bra\Delta\ket$ increases in the case of $d_{3/2}$ with $\lambda=-1.0$
MeV.

\begin{figure}[htbp]
\includegraphics[scale=2,angle=0]{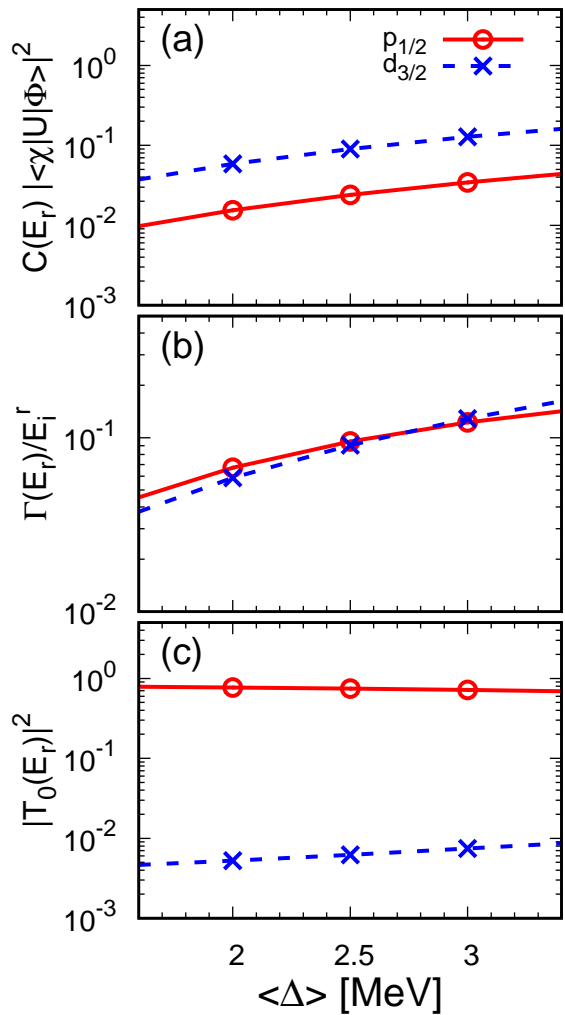}
\caption{(Color online) The pairing gap $\bra\Delta\ket$ dependence for the numerator
  of Eq.(\ref{q2er}), the scaled width $\Gamma/E_i^r$ and $|T_0|^2$ are shown in panels
  (a), (b) and (c) respectively. The red solid and blue dashed curves represent $p_{1/2}$
  with $\lambda=-8.0$ MeV and $d_{3/2}$ with $\lambda=-1.0$ MeV, respectively.
  A coefficient used for (a) is given by $C(E_r)=\left(\frac{2m}{\hbar^2}\right)^2\frac{1}{k_1(E_r)}$.}
\label{fig3}
\end{figure}

In order to clarify the pairing effect on $q_{lj}(E_r)$, we analyzed the
pairing gap $\bra\Delta\ket$ dependence of Eq.(\ref{q2er}) in Fig.~\ref{fig3}.
The numerator of Eq.(\ref{q2er}), the scaled width and $|T_0(E_r)|^2$
are plotted as a function of the pairing gap $\bra\Delta\ket$ in the panel (a),
(b) and (c), respectively. The red solid and blue dashed curves represent
$p_{1/2}$ and $d_{3/2}$ respectively. The numerator of Eq.(\ref{q2er}) and
the scaled width increase as the pairing gap increases. The numerator of
Eq.(\ref{q2er}) for $d_{3/2}$ (with $\lambda=-1.0$ MeV) is larger than
the one for $p_{1/2}$ (with $\lambda=-8.0$ MeV). This indicates that $d_{3/2}$
is more sensitive to the pairing than $p_{1/2}$ because $d_{3/2}$ is closer
to the Fermi energy $\lambda$ than $p_{1/2}$.
In the panel (b), the scaled widths show almost the same values and dependence
on the pairing gap. The reason is rather trivial. As shown by Eq.(\ref{defgm}),
the width is expressed by the square of the coupling strength by pairing
between the HF hole state and continuum. Therefore, the width should be similar
value with the same pairing gap. Also, both quasiparticle resonances appear
at the similar incident energies as shown in Table~\ref{table1}.
The $|T_0(E_r)|^2$ values for $p_{1/2}$ are much larger than the one for $d_{3/2}$.
The difference of the $q_{lj}^2(E_r)$ between $p_{1/2}$ and $d_{3/2}$ is due
to the difference of the ratios of the transition probability to the
``{\it modified quasi-hole}'' state and to the HF continuum between
$p_{1/2}$ and $d_{3/2}$.

\section{Conclusion}

The small absolute value of $q_{lj}(E_r)$ for $p_{1/2}$ with $\lambda=-8.0$ MeV
is due to the large value of the transition probability to the HF continuum and
the small transition probability to the ``{\it modified quasi-hole}'' state
because of the small pairing effect for the deep-lying hole state.
However, the small absolute value of $q_{lj}(E_r)$ causes the asymmetric shape of
the partial cross section of the neutron elastic scattering which is known as a
typical sign of the Fano effect.

On the other hand, the large absolute value of $q_{lj}(E_r)$ for $d_{3/2}$ with
$\lambda=-1.0$ MeV is due to the small value of the transition probability to
the HF continuum and the larger transition probability to the
``{\it modified quasi-hole}'' state. The hole $d_{3/2}$ state with $\lambda=-1.0$ MeV
is much closer to the Fermi energy than $p_{1/2}$ with $\lambda=-8.0$ MeV at
the zero pairing limit, $d_{3/2}$ state with $\lambda=-1.0$ MeV is more sensitive
to the pairing effect. This is the reason of the larger transition probability to the
``{\it modified quasi-hole}'' state. The shape of the cross section for
the quasiparticle resonance becomes the shape of the Breit-Wigner formula with
the large $q_{lj}(E_r)$. More Breit-Wigner type resonances are, therefore expected
to be observed in the neutron elastic scattering cross section on the neutron-rich
open-shell nuclei.

In this study, we show the possibility of the discussion of the pairing correlation
and the single-particle level structure of the target nucleus of the neutron elastic
scattering in terms of the quasiparticle resonance and the Fano effect. However,
the channel-coupling effect is known as the origin of many of sharp resonances observed
in the neutron elastic scattering cross section as clarified by the R-matrix
analysis~\cite{rmatrix} and the cPVC calculation~\cite{mizuyama}.
The Fano effect due to the channel-coupling is also expected.
The channel-coupling and pairing correlation need to be taken into account
for the actual analysis of the experimental data.

\section{Acknowledgments}
This work is funded by Vietnam National Foundation for Science and Technology
Development (NAFOSTED) under grant number “103.04-2018.303”.

\end{document}